*Article*

# Localized surface plasmon enhanced InGaN-based light-emitting diodes with polygonal Au/Al nanoparticles

Xiaokang Mao[1], Zhiguo Yu[2], Peng Chen[1,*], Jing Zhou[1], Yuyin Li[1], Jianbo Feng[1], Zili Xie[1], Hong Zhao[1], Dunjun Chen[1], Xiangqian Xiu[1], Ke Wang[1], Yi Shi[1], Rong Zhang[1], Youdou Zheng[1]

[1] Jiangsu Provincial Key Laboratory of Advanced Photonic and Electronic Materials and School of Electronic Science and Engineering, Nanjing University, Nanjing, 210093, China
[2] Institute of Semiconductors, Chinese Academy of Sciences, Beijing, 100083, China

**Abstract:** Localized surface plasmons (LSPs) have played a significant role in improving the light emission efficiency of light emitting diodes (LEDs). In this report, polygonal nanoholes have been fabricated in the p-GaN layer of InGaN-based LEDs by using Ni nanoporous film as the etching mask, and then Au/Al metal nanoparticles are embedded in the nanoholes to form the LSP structure. The coupling between the LSP and the LED has been clearly observed. The results show that the light output of the LEDs has been increased by 46% at higher current injection condition, and together with a shift of the gain peak position to the LSP peak resonance energy. As the coupling distance is decreased from 60 nm to 30 nm, the maximum enhancement factor increases to 2.38. The above results indicate that the LSP from the polygonal metal nanoparticles is a kind of very promising structure to enhance the lighting performance of the InGaN-based LEDs.

**Acknowledgments:** This work is supported by National Nature Science Foundation of China (12074182), Collaborative Innovation Center of Solid-State Lighting and Energy-saving Electronics, Open Fund of the State Key Laboratory on Integrated Optoelectronics (IOSKL2017KF03) and the Research and Development Funds from State Grid Shandong Electric Power Company and Electric Power Research Institute.

* Email: pchen@nju.edu.cn

This work has ever been successfully submitted to *Crystals* for possible publication in Feb. 2021, and the manuscript ID is crystals-1129650 .

## 1. Introduction

In the past two decades, GaN-based blue light-emitting diodes (LEDs) have attracted extensive attention and achieved great success [1, 2]. High-Power LEDs have been used for Solid-State Lighting and GaN-based blue laser diodes used for high-density data storage and laser display [3, 4]. The internal quantum efficiency of blue LEDs can reach over 70% [3]. However, due to the low quality of indium gallium nitride (InGaN) materials with high indium (In) content [4] and the aggravation of polarization [5], the internal quantum efficiency of the LEDs with high In content or at high current injection is becoming low, for example, it can be decreased from over 60% to below 50% at high current of 1000mA[3] . Among the ways to improve the internal quantum efficiency of InGaN-based LEDs, coupling with localized surface plasmons (LSPs) is a very promising method [6–12]. When the quantum wells (QWs) is placed in the evanescent field of LSPs, the QWs can be strongly coupled with the LSPs once the energy of the exciton in the quantum well matches the resonance energy of the LSPs. Because of the extremely high density of states in localized surface plasmon resonance, energy can be transferred from QWs to surface plasmons (SPs) with very high efficiency, more than 90% [13]. Here, we can say there may also be another energy induction mechanism, that is improving the radiative recombination of the excitons locally from the LSP resonance, which still is an open topic. Overall, the spontaneous radiation efficiency of the QWs is improved, and then the internal quantum efficiency is improved. Generally, the penetration depth of SPs in normal semiconductors is only tens of nanometers [6], but the thickness of the p-type layer of conventional LEDs can be about 200 nm [8]. So, there must be some special design to achieve the coupling between SPs and the QWs. So far, people have developed some device structures that realize LSPs-enhanced LEDs. Yang et al. achieved QW-SP coupling by depositing Ag nanoparticles on p-GaN with a thickness of 80 nm [7]. Cho et al. realized SPs-enhanced LED by embedding metal particles in the p-GaN layer near the QWs region during the growth process [9,10]. Cheng-Hsueh Lu reported another coupling structure, i.e., they fabricated p-type GaN into a porous structure, and then evaporated a layer of silver particles in the holes to produce SPs. This method can avoid additional metal deposition process in the growth process [11]. Currently, most of results are based on photoluminance (PL) measurement, and it is not easy to achieve electroluminescence of SPs-enhanced LEDs by using an effective LSP structure.

In this study, LSP-enhanced LEDs were fabricated and analyzed by using polygonal Al/Au nanoparticles as the LSP structure. The nanoporous nickel was used as an etching template to realize the polygonal nanoholes in the p-type GaN of InGaN-based LEDs. Al/Au bilayers were deposited at the bottom of the nanoholes to couple with the QWs within tens of nanometers. The polygonal metal nanoparticles have sharp corners which can even increase the plasmon effect by the Tip Effect. The emission characteristics of the LEDs have been measured finally.

## 2. Materials and Methods

The LED wafer used in this study was grown on a c-plane sapphire substrate by metal organic chemical vapor deposition (MOCVD). The device structure is shown in Figure 1, including 4-μm undoped and Si-doped GaN, five-period InGaN/GaN QWs, and 200-nm p-type GaN. For the fabrication of the polygonal nanoholes, a 6-nm Ni was deposited on the LED surface by electron beam evaporation (EBE), following with rapid thermal annealing (RTA) to form nanoporous mask with proper temperature and duration. Then the LED wafer was etched with the nanoporous mask by inductively coupled plasma etching (ICP). In order to study the influence of different coupling distances, the etching depth was controlled at 140 nm and 170 nm respectively. By this way, LED chips with three structures on the p-type GaN layer are obtained, one is the unetched planar p-GaN (plane-LED), another two are nanoporous p-GaN with a depth of 140 nm (pore140-LED) and with a depth of 170 nm (pore170-LED).

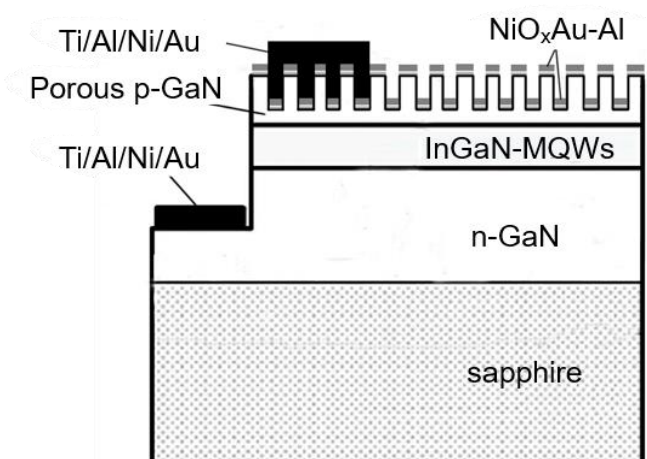

**Figure 1.** The structure of the Al/Au metal particle LSP enhanced LED.

After etching, the remaining Ni film was washed away with dilute nitric acid, and then the entire LED chip was prepared for depositing electrode and LSP metal. The n-type electrode of Ti/Al/Ni/Au (30/170/50/150 nm) was deposited by electron beam evaporation. Then Ni/Au (5/5 nm) was deposited on the porous p-GaN layer as a current spreading layer and annealed at 500℃ in air for 1 min. As Ni will be oxidized during annealing, the metal at the bottom of the hole was 5-nm Au only [14]. Then, 20-nm Al was deposited at last, so that the metal at the bottom of the hole was Au/Al (5/20 nm) double layer structure. Finally, Ti/Al/Ni/Au (30/170/50/150 nm) was deposited as p-type electrodes.

The p-type GaN surface nanoporous structure was characterized by scanning electron microscope (SEM) and atomic force microscopy (AFM). For LED devices with different etching depths of p-GaN, electroluminescence (EL) tests were performed from the top surface and back side of the LED devices, respectively.

## 3. Results and Discussion

### 3.1. Characterization of porous mask and p-GaN surface

Figure 2 shows the surface morphology of the nanoporous structure realized after-annealing Ni on the p-GaN surface. It can be seen from Figure 2 (a) that the nanoholes formed after annealing of the Ni film, which present a polygonal morphology. The polygonal nanoholes have the common feature that the corners are 60° or 120° [15]. The average hole size is about 200±50 nm and the spacing between holes is about 400±100 nm. Figure 2 (b) is an AFM image of the etched p-GaN surface. After comparison with Figure 2 (a), it can be seen that hole patterns have been successfully transferred from the Ni nanoporous mask to the p-GaN layer. Figure 2 (c) shows the cross-sectional profile from the black line in Figure 2 (b). In this sample, the hole depth is basically maintained at 170 nm. Considering that the thickness of p-GaN is 200 nm, the distance between the bottom of the pore170-LED and the active area is only 30 nm, which is less than the penetration depth of SP. The penetration depth (Z) here can be approximately given by,

$$Z = \lambda/2\pi [\,(\varepsilon'_{GaN} - \varepsilon'_{metal})/{\varepsilon'_{metal}}^2\,]^{1/2} \qquad (1)$$

where $\varepsilon'_{GaN}$ and $\varepsilon'_{metal}$ are the real parts of the dielectric constants of GaN and metal. Based on this equation, Z can be calculated as 77 nm for Al and 33 nm for Au, respectively[6]. Thus, in our samples, Al will give more important role by considering the metal thickness and its Z value.

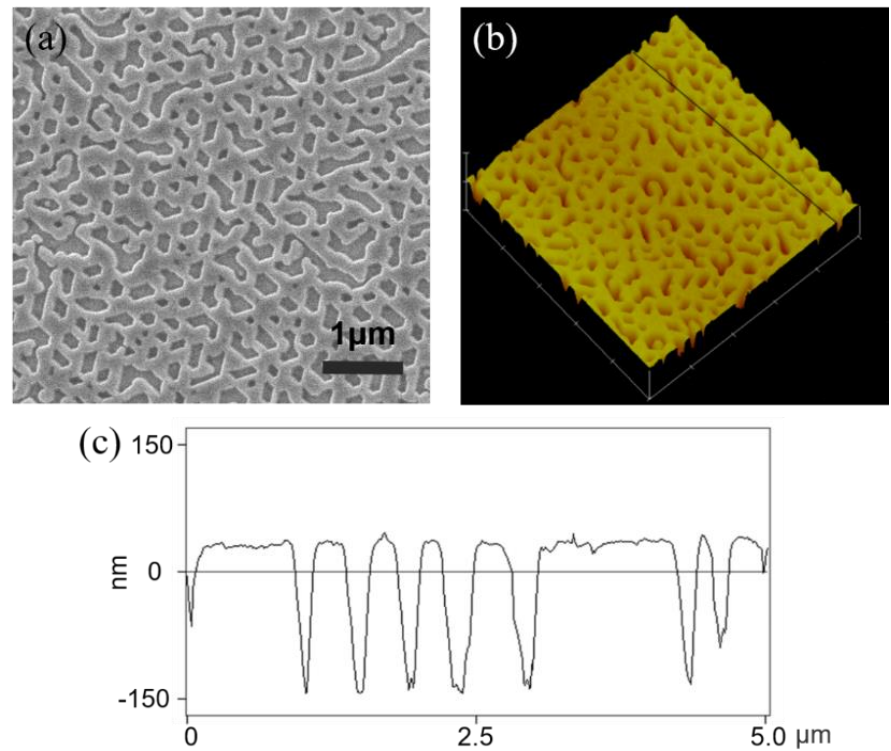

**Figure 2.** The morphology of the nanoporous structure on the p-type GaN surface: (a) SEM image of the Ni porous template, (b) AFM image of the p-type GaN surface nanoporous structure of pore170-LED, (c) the cross-sectional morphology at the line in (b).

*3.2. The I-V and Optical Characteristics of LSP Enhanced LEDs*

Figure 3 shows the I-V characteristic curves of the three structures. In the case of low voltage (<3V), as the depth of the hole increases, the forward currents of the LSP-coupled LEDs are larger than the plane LED, which is mainly caused by the leakage current due to the etching damages. When the voltage is above 3V, as the depth of the hole increases, the forward currents of the LSP-coupled LEDs are lower than the plane LED, which is mainly caused by the higher resistance due to the smaller effective injection area of the p electrode.

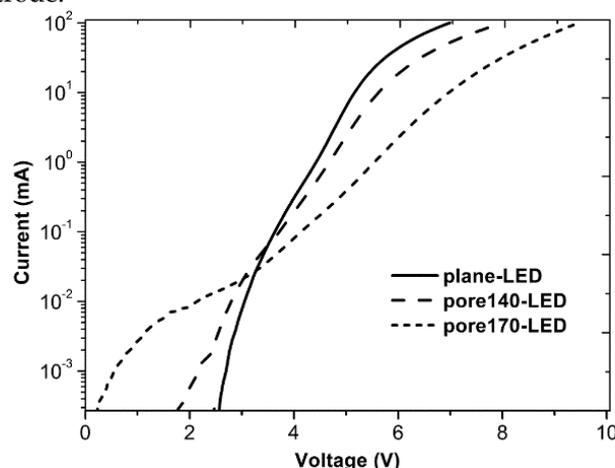

**Figure.3** I-V curves of plane-LED, pore140-LED and pore170-LED

It has been known that the surface roughness caused by the nanoporous p-type surface can increase the light extraction efficiency of the top surface, which increases the light scattering at the top surface and then reduces the total reflection. [16]. At the same time, the surface-plasmon coupling can also increase the emission intensity, but not limited on the top surface [17]. Therefore, it is necessary to perform the EL spectrum test from the top and back side of the LED devices at the same time.

Figure 4 (a) shows the EL spectra of the plane-LED, pore140-LED and pore170-LED at a current of 20 mA, which are already the sum of the top and bottom emission spectra. The inset shows the absorption spectrum of Au/Al particles embedded in the p-GaN hole, and it indicates that the peak of the Au/Al LSPR is at 523 nm in the extinction spectrum [18]. From Figure 4(a), a redshift can be seen for the devices with LSP coupling, and the redshift is increased with the increase of the hole depth, i.e. the stronger coupling. As the hole depth increases, the spectrum peak of the LED gradually shows the redshift from 435.9 nm of plane-LED to 439.9 nm of pore140-LED and 441.6 nm of pore170-LED respectively.

This redshift cannot be explained by the porous structure only. The porous structure may induce the stress releasing and an increase in current density because of a reduction in effective injection area. But, both stress releasing [15] and the increase of current density [19] can only cause a blueshift in typical InGaN-based LEDs.Therefore, the strain relaxation and higher current density will cause the blue shift of the emission peak wavelength, which is contrary to our experimental observation. So, the surface roughing may partially contribute to the increase of the light output, but there are other reasons. On the other hand, by comparing the absorption spectrum with the EL spectrum, it can be found that the EL change satisfies the typical phenomenon of SPs coupling, that is, the peak of the EL spectrum moves to the direction of LSPR peak. As metal particles get closer to the MQWs, the luminescence peak shift amplitude increases, further to the LSPR peak [6, 20]. Therefore, it can be concluded that the main reason for the EL spectrum redshift in the coupled LED is the coupling of Au/Al particles and the MQWs.

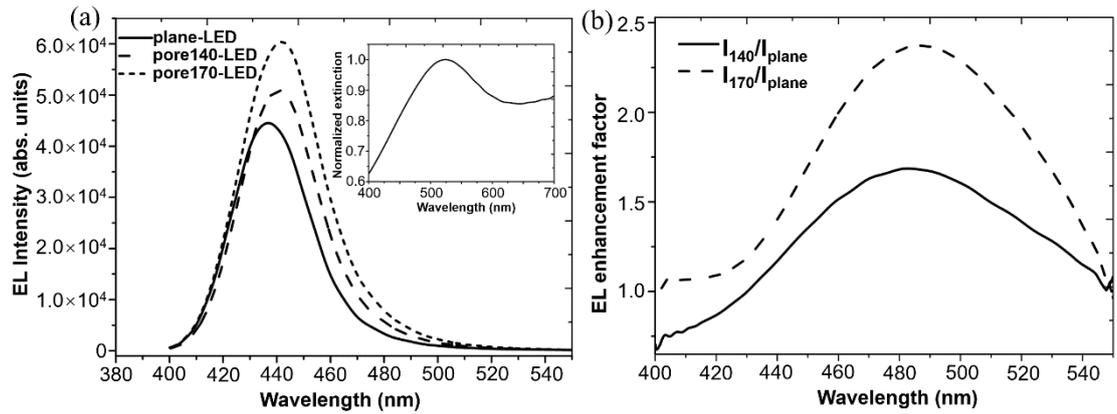

**Figure 4.** (a) The EL spectrum of the total light output power by the sum of the front and back light; the inset is the absorption spectrum of Au/Al particles. (b) EL enhancement factor curve of pore140-LED and pore170-LED.

Besides the redshift, the emission intensity has also been modulated by the surface structure. The emission intensity, $I_{plane}$, is for the plane-LED, and similarly $I_{140}$ is for pore140-LED, $I_{170}$ is for pore170-LED. By calculating the ratio between the coupled samples and the plane-LED, we can obtain the wavelength-depended EL enhancement factor curve, $I_{140}/I_{plane}$ and $I_{170}/I_{plane}$, as shown in Figure 4(b). For pore140-LED, the maximum value of EL enhancement factor is 1.69, and the corresponding wavelength is 482.1nm. As the hole depth increases from 140 nm to 170 nm, the maximum enhancement factor increases to 2.38, and the peak wavelength shifts to 486.1 nm. Considering the exponential decay characteristics of the LSP evanescent field, the increase of EL enhancement factor and the red shift of the enhancement peak with the increase of the hole depth can be explained by the stronger LSP coupling with the closer distance. As the distance between Au/Al particles and MQWs decreases from 60 nm to 30 nm, the QW-SP coupling becomes stronger.

The strong QW-SP coupling can result in two different energy transfer routes. One is transfer of the exciton energy to the LSP and realizing photon emission, which is common mode reported and presents a fast decay of the emission. Here, we propose the second energy transfer route, that is the LSP can modify the recombination behavior of the excitons that distribute within the LSP evanescent field, which can result in a higher emission efficiency by a faster spontaneous emission rate or a longer lifetime of the excitons in QWs. Here, it has to be pointed that the ostensive decay time from a time resolved PL cannot simply relate to the carrier spontaneous emission rate. An obvious counterexample is that a shorter decay time can be caused by defects, which may capture and deplete the carriers faster. In the absence of an electric field, longer carrier lifetimes generally give higher spontaneous recombination efficiency, which means fewer loss channels for the carriers. The radiative recombination dynamics process the second energy transfer route needs further investigation and our initial time resolved PL experiment shows that a longer lifetime has been observed in the sample of pore170-LED and pore140-LED, which will be published later. The increase of carrier lifetime through our LSP coupling is an obvious sign. One of possible reason may be from the polygonal Au/Al nanoparticles that have sharp corners and can even increase the plasmon effect by the Tip Effect. It also has to be pointed that the EL enhancement is not only related to the density of states of LSP, but also affected by factors such as the density of states of the excitons and the internal quantum efficiency of the quantum well itself [6, 21, 22]. This may contribute to the differences between the position of maximum enhancement factor and resonance peak position of the LSP.

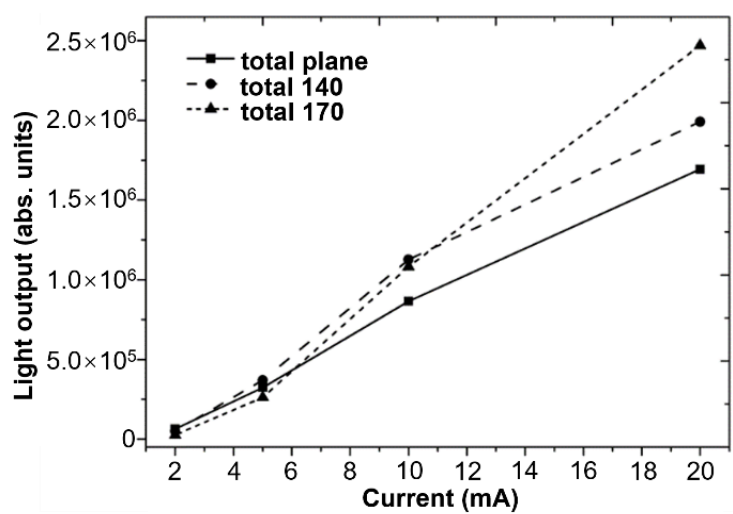

**Figure 5.** Light output power curve of plane-LED, pore140-LED and pore170-LED under different working currents

In order to further investigate the properties of LSP-enhanced LEDs, the relationship between the EL spectrum integrated intensity and the current was tested for all three samples, as shown in Figure 5. In the case of low current (I=2 mA), the integrated EL intensity of the plane-LED is higher than those of the pore140-LED and the pore170-LED. With the increase of current, the integrated EL intensities of the pore170-LED and the pore140-LED is higher than that of the plane-LED gradually. When the current is 20 mA, the integral intensities of the pore140-LED and the pore170-LED are 18% and 46% higher than that of the plane-LED respectively. Considering the above analysis, it can be seen that with the working current of 20 mA, the stronger enhancement can be obtained in the stronger coupled sample. In the case of low current, the integrated intensity of the LSP-coupled LEDs is smaller than that of the planar device, which may be due to the large leakage current [23] of the porous structure caused by the lattice damage from the dry etching [24]. However, as the working current increases, the leakage current will reach saturation and the total current mainly comes from the diffusion current. This means that more current can contribute to the recombination in the active area under high current injection, so the effect of LSP enhancement is prominent now. It shows a special meaning that the stronger enhancement is observed by the effect of LSP coupling at higher working current. It is well known that the InGaN-based LEDs always encounter a serious problem, i.e. efficiency droop at high current injection. So, from this study, the LSP coupling provides a very promising solution for high-efficiency InGaN-based LEDs with the second energy transfer route.

## 4. Conclusions

From this study, nanoporous nickel can be used as an etching mask to prepare nanoporous p-GaN structured LEDs. The LSP from the polygonal Au/Al nanoparticles can effectively couple with the QWs in the LEDs. Through the coupling between the LSP and QWs, the light output power of the LSP-coupled LED shows an increase of 46%. One new mechanism of LSP coupling to enhance output of the LED is proposed and the LSP coupling provides a very promising solution for improving the light output of InGaN-based LEDs.

**Author Contributions:** Xiaokang Mao and Zhiguo Yu contributed equally to this work. Zhiguo Yu performed the fabrication process and optical measurements. Peng Chen conceived the idea of the whole experiments and analyzed the data. All authors discussed the results.

**Author Information:** The authors declare no competing financial interests.